# Multi-agent Searching System for Medical Information

MARIYA EVTIMOVA- GARDAIR
FKSU
Technical University of Sofia
Technical University of Sofia, bul.Kliment Ohridski 8, Sofia
BULGARIA
mevtimova@tu-sofia.bg

*Abstract:* - In the paper is proposed a model of multi-agent security system for searching a medical information in Internet. The advantages when using mobile agent is described, so that to perform searching in Internet. Nowadays, multi- agent systems found their application into distribution of decisions. For modeling the proposed multi- agent medical system is used JADE. Finally, the results when using mobile agent are generated that could reflect performance when working with BIG DATA. The proposed system is having also relatively high precision 96% .

## 1 Introduction

Nowadays, it is observed a growing interest in studying multi- agent systems for searching of information. In the most papers that describe multi-agent systems, is observed the interest with coordination of the agent behavior and distributing or unifying decisions that make them convenient to work with big data. Furthermore, multi- agent systems are one developed direction that is applied when providing theoretical and laboratory research to searching systems. Agents may include characteristics, such as autonomy, reactivity, proactivity, and social capabilities. These features concern their behavior and interaction with the surrounding environment, which is important for their application in information retrieval systems. For the agent, the environment is everything external, but have continuous connectivity and interaction with it.

Searching systems that are on the market have a number of disadvantages and do not always meet the need for medical information of the user that have health problem. An artificial intelligence is proposed solution for efficient searching system in internet.

The need for developing a system for searching medical information in internet is derived from the statistic that a lot of people and also medical specialists are searching medical information in internet. The results derived from the system could be crucial for diagnostics of a certain disease. So that, implementation of security to the proposed system is necessary to improve the quality of the proposed searched system for medical information. Therefore, to improve the system it is necessary to implement data privacy preserving analysis and also data need to be secured before data analysis.

From the other side, as a part of Distributed Artificial Intelligence (DAI) the multi- agent systems and autonomous agents give a new method for analyzing, designing and implementing complex applications. In recent years, the majority of applications optimize their goals with distributed tasks between autonomous agents [1]. This assure intelligent agents to return to the users accurate and relevant (up-to- date) results. This is a consequence of the ability of intelligent agents to help the user find and filter information on the web. That give advantage when using intelligent agents into search systems, which enable the system to work high precision and can learn from their environment in order to reproduce accurate and relevant results [2].

The requirement to continuously improve the quality of the searching systems implies the standard for all new information systems to evaluate the quality of the returned results and compare their results with the other created search engines.

## 2 Description of the proposed searching system for medical information

### 2.1 Conceptual schema of the multi- agent security searching system for medical information

The conceptual schema in Fig.1 define a system that can retrieve relevant medical information that correspond to the customer health problem [3][4][5]. That conceptual schema can be divided into three basic parts. The first part consist of security log- in authentication of each user. The second part include definition of the query that is requested from the user. And the third part include the mechanism of searching information from Internet sources with mobile coordinating agent. Agents represent software parts that have their specific goals and have the possibility to function autonomously in a certain environment and is able to make contact with another agent or group of agents. Using agent based programming have advantages that make him a good choice for an information system [6]. The agent have the ability to interact with the environment and this can help to perform dynamic update of the user preference. In addition the system requires a lot of interconnections between the user and the system to define the query and consecutively to return quality results. Furthermore, the agents are preferred to be used for effective communication and also their level of abstraction. In the proposed conceptual schema the user first create a profile where he can add common information, medical information and health condition and then the user can enter the text request into the interface. Then the user request is modified concerning the content of the user profile, so that the returned results to become personalized.

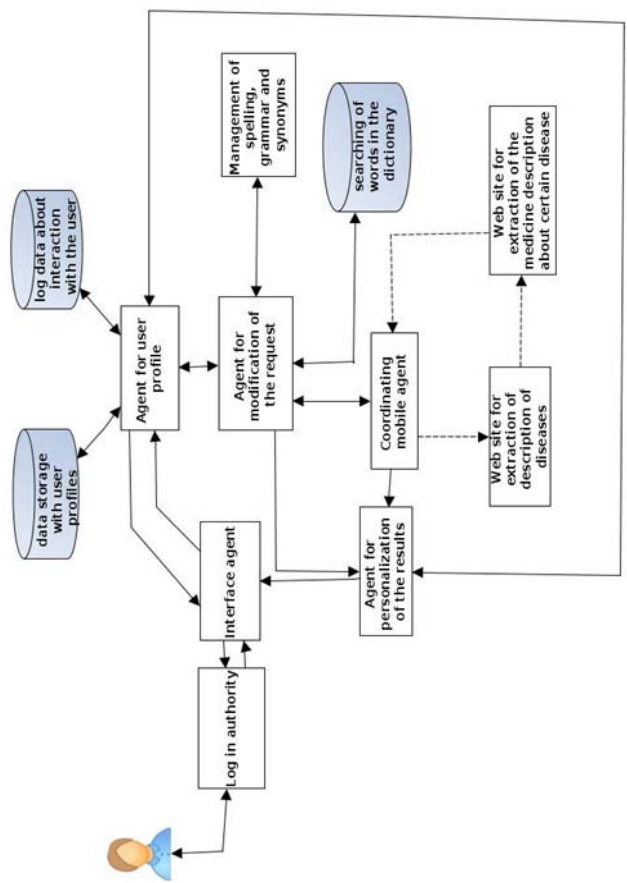

Fig.1: Conceptual schema of the proposed searching system

The conceptual scheme consists of five agents and security module with log- in authority. The interface agent that interact with the user, agent for user profile that captures and maintains user preferences, agent for modification of the query, agent for returned results personalization and a mobile agent that searches information on the websites. In Fig.1 it is shown the conceptual scheme that is proposed in the article.

1. Description of the module with log- in authority

In Fig.1 as "log- in authority" is presented the part of the searching system with the security mechanism. The need of adding security into the searching system is derived from the web system's vulnerabilities that can give deviation when making personal medical analysis. Furthermore, it is necessary to perform a data privacy preserving analysis and also implementation of data security before continuing with data analysis. The mechanism that is proposed for the defined system security is with IP address and user Identification number.

The security mechanism applied is based on following principles:

1. No one user can be identified after using the system with the medical data from a user in the real world
2. Each web site that is used from the system need to provide level of identification assurance as high as the entity that provide the medical data.
3. For every searching of information in the system for a user is used different key to index a user's records that prevent the traceability of the medical data

2.Interface agent
The interface agent include the interaction with the user after login to retrieve the user input and display the retrieved results. The input data can be the preferences principally entered from the user. These request that represent their interest and can be accepted from the user profile agent so that to update the user profile. The input data could be the user requests that are transferred to the agent that define the query and create the modification of the query. Also, the interface agent formulates the personalized user results. User interconnections can be explicit or implicit. Explicit interactions can be expressed when asking the user about his feedback on the results and implicit interactions can be expressed when monitoring the user behavior of the results.

3.User profile agent
Agent of the user profile have functionality to manage the profile. The user is asked to fill in a form that correspond to the user preferences. Beside many users do not like to fill in that form, because it takes time, it helps to increase the quality of the returned results that correspond to the personalization.

4.Agent for modification of the results
Agent for modification of the query will interconnect to the user, so that to review the user query and append any luck of information or spelling adjustment. After receiving the user request, the agent switch to appropriate language, so that to propose relative syntax and processing. The query request is divided into words, then is performed spellcheck to verify spelling of the words and after that is applied synonyms manager service to provide synonyms for each term. Thereafter, the query is classified using the word search service in the dictionary. The dictionary for searching words is a predefined term storage that helps to identify stop words and relationship between the terms. This is a multilingual vocabulary where there is a list of terms in each language that is used for searching and matching the terms of the query. The next step is to filter the unnecessary words using a word search dictionary that has a list of redundant words that are defined, such as: between, do, on and etc. Other terms are then identified as possible links between these terms, using a word search dictionary that matches predefined terms. The agent also gets the synonyms of the spell checker terms and the synonym manager to link it to the terms of the user request. In the end, the information is submitted to the mobile agent to find the relevant terms in the request on the relevant web site with information.

The information gathered from the mobile agent sends the data to the agent to personalize the results. If the user request does not match, then the agent review the request using the information from the user profile.

Finally, the application annotation is produced and sent to the coordinator agent

5.Agent for personalization of the results
The agent responsible for personalization of the results improve the returned results to the user. Firstly, it receives an annotated request from the agent for modification of the query and then use the appropriate algorithm to extract the appropriate result of the requested query. In this agent, results are further processed when released from conflicts, assembling similar results, ranking and sorting after receiving the user preferences taken by the profile agent. Finally, the results are personalized and sent to the interface agent, so that to be displayed to the end user.

6.Mobile agent
Mobile agent have to get the requested query from the user. The agent is able to find the relevant information on the web about a certain request from the user and then that information will be send to the agent for request personalization. Defined as a goal of the mobile agent is to get the updated information about a user request and to search for relevant information from websites. After an agent for modification of the request receives a message, it search to send the request to the coordinating mobile agent. The coordinated mobile agent retrieves the desired categories from the text of the request and searches all over the locations located in those categories. Once they have been discovered, the mobile agent move from one place (web site) to another. As defined each place has a corresponding web site, so when the agent arrives at the web site, it starts collecting the data from the relevant page and gathers them in a database and then move to the next place. This action is repeated until all places are visited and then the mobile coordinating agent informs the agent for modification of the query that the data has been collected.

## 2.2 Searching with the proposed searching system of information on the web when using coordinating mobile agent

Functionality of the agent includes learning, planning and searching for the current information in Internet. Collecting information is a difficult process that depend from the type of the collected information, but researchers are trying to improve current methods or even find new ones. The present coordinating mobile agent accept the user request from the agent of the modification of the query, then it finds the required category, and transfers the request to the appropriate web agent that should be used, if the agent was not mobile. This diagram is shown in Fig.2. Fig.3 show a schema with coordinating mobile agent that moves through all locations in these categories. When the agent arrive in a definite place the mobile agent extract all the requested information from the relevant web page. The scheme with the mobile coordinator agent is shown in Fig.3.

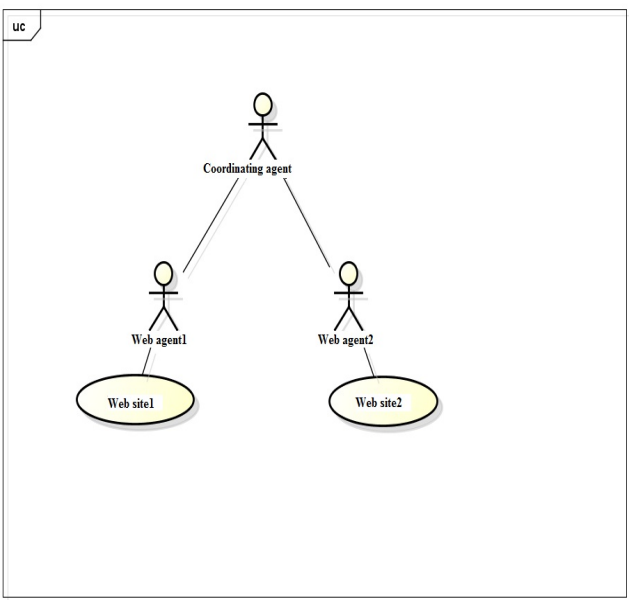

Fig.2: Collecting information with static agent

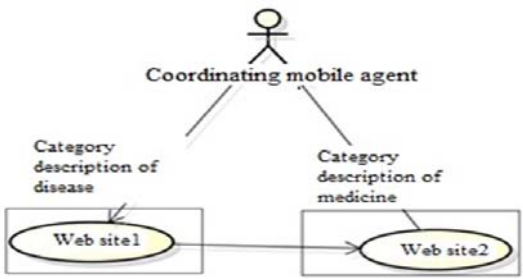

Fig. 3: Collecting information with mobile agent

Using a coordinating mobile agent saves the creation of web agents when making the system model simpler.

## 2.3 Presentation of the static and mobile agent when searching for information on the Internet

The sniffer agent in JADE allows real- time monitoring of how agents interact with each other to solve a problem. This sequence is described in Fig.4, as it present the search of information on the Internet when using static agent.

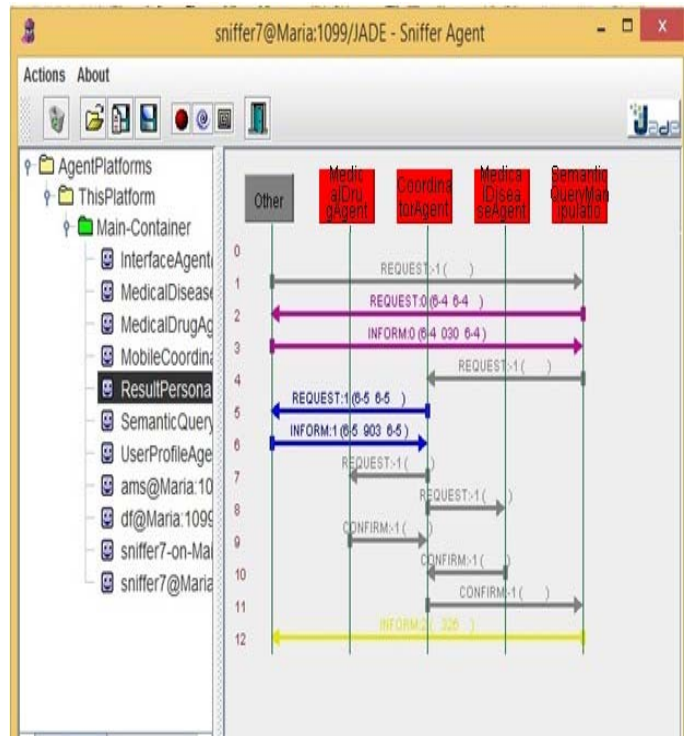

Fig. 4: The Sniffer agent tracks the sequence of the communication between the agents when using static agent

When using a mobile agent, its functionality replaces that of the coordinating agent and web agents, making the scheme simpler. As it could be observed in Fig.4, when the agent for modification of the request accept a request from the user. It searches an agent that offer the coordination service through the request message and receives a response by informing the communication activity.

After that it sends a request to a coordinating agent who will ask for the agents located in the desired categories. And then it sends a request to a coordinating agent who will ask for the agents located in the desired categories. Once found, the coordinator will send requests to those funds awaiting their confirmation. When all of them get confirmation that they will send a confirmation message to the agent for modification of the request, then they have the possibility to learn when the request is completed.

## 3 Implementation of the searching system in Internet

The application is implemented using Java Technologies such as JADE and HtmlUnit. The JADE function for agent execution is used. During startup, each agent registers at Agent Management System (AMS) and then registers his service with Directory Facilitator (DF) yellow pages. The services are distinguished by their description so that each web agent is registered with the name of his category, the agents register with their services. In this way, agents can easily find interaction and communicate to achieve their goals.

When registering agent services, "type" is the type of the agent service and "name" is the name of the service. The mobile coordinating agent can be registered as "coordinator". The mobile coordinating agent replaces the role of the web agent (when using a static coordinating agent) and retrieves information from a specific web page. This process is done using the getResults() method of the class of the corresponding web page.

This method uses HtmlUnit class libraries to fill the search form of the page, click the send button and then analyze the resulting HTML page to collect the data. Algorithm. 1 shows an algorithm written in the JAVA language using the JADE platform. JADE demonstrates the algorithm that searches for the www.medicine.com disease description site.

The collection of data is presented using HtmlUnit with the following algorithm:

ALGORITHM 1: Iterative Algorithm

get final results from the Web page

get information from the HTML web form:

**getElement** from HTML form as **disease search**

**setAttribute** value for the **getElement** to **disease**

**clickButton** to submit from the HTML page

**end**

In the same way, information about the medicine for the relevant disease is also collected from http://www.drugbank.ca/.

The coordinating mobile agent combines the role of the coordinating static agent and web agent. The coordinating mobile agent first find all available places in the platform through GetAvailableLocations behavior by submitting a request to the Agent Management Service.

After receiving the request, he will filter the list of places and choose only those who need to use the method for filtering the locations, then start moving to the websites using the move() method. The method that is responsible after the movement tells the agent what to do after the agent arrives.

Depending on the place where it arrives, it will begin collecting data from the web site. This is repeated until all places are visited and he informs the agent handling the request that he has to complete the task.

## 4 Evaluation of the use of mobile agents when searching for information on the Internet

### 4.1 Evaluation concerning mobile and static agent usage

In order to justify the use of mobile agent when searching information, a short comparative analysis with results is presented. This is shown in Table 1. These results are obtained by performing the searching of data in Internet with mobile or static agent. For system testing, a computer with the following system configuration is used: Intel Core i5-5200 CPU @2.20GHz, 4GB RAM. A detailed description of the results obtained is presented in the following paragraphs.

Table 1: Comparative results when searching information with static and mobile agent.

|  | Searching with static agent | Searching with mobile agent |
|---|---|---|
| Times for response | 80524ms | 75123ms |

The total time it takes to get a response from static agent architecture will be the maximum time it takes for a web agent to collect all the data from the web page plus the time for agent communication. Compared to this, the total time required for mobile agent architecture will be the sum of the times needed to collect the data from each web page. The time required for communication is reduced in mobile agent architecture because of the reduced number of agents involved.

At first glance, it seems obvious that the system works better with static agents, but actually when making analysis it observed different results. The larger number of agents running the system at the same time on the same platform, even those waiting, presents a large number of computer processes that need to be handled, which makes it harder for agents to work and filter web pages when looking for the desired information. This aspect makes the period of time used to collect data grow, that result to a slower system. Based on these results, mobile agent architecture is more preferred because of the need for resource management. Another advantage that can be added is the security of the communication channel communicated by the agents. The test results can be changed when the application of a different system configuration is started at another speed of the Internet connection. So that, the mobile agent architecture approach has advantages in the current context: the way data is collected in combination with the system, but when information gathering is improved on the side of the information, there is a good chance that a static agent approach is better.

## 4.2 Evaluation the results when testing the system with mobile agent

The proposed system with mobile agent is tested with 225 different requests concerning each different category of the symptoms like: abdominal symptom, cardiovascular system symptom, digestive system symptom, head and neck symptom, hemic and immune system, musculosceleton system symptom, nervous system symptom, neurological and physiological symptom, nutrition, metabolism and development symptom, reproductive system symptom, respiratory and chest symptom, skin and intergumentary tissue symptom and urinary system symptom.

The results from the different category are assumed and give the following estimation of the system performance: Precision 96%, Recall 91% and F-measure 93%.

## 5 Conclusion

In the paper is described a medical security web system for searching of information in Internet that mobile agent. Nowadays more and more people professional and not professional are searching medical information in Internet and that system will help them to select the appropriate information. It is performed the evaluation of the system and the results define the system with reasonable high quality comparing to other medical systems for searching of information.

## 6 ACKNOWLEDGMENTS

The research that the results are presented this publication is funded by the Internal Grant no: 191ПР0006-09 of Technical University- Sofia 2019.